# Old equations for a new system:

# A possible use of Navier-Stokes equations to model the circulation of spikes in the nervous system

**Daniela Sabrina Andres**

*Department of Physiology, Medicine School, University of Buenos Aires*

*CONICET*

*Institute for Neurological Research Raúl Carrea, Movement Disorders Section, Neuroscience Department, FLENI*

*Montañeses 2325, C1428AQK, Buenos Aires, Argentina*

*danielaandres@conicet.gov.ar*

**Abstract**

In the present work we discuss a possible application of Navier-Stokes-based models to the quantitative description of the circulation of nervous impulses throughout the nervous system. Velocity and momentum of spikes are not usually considered when modeling information transmission in the nervous system, however in previous works we have shown that the discharge from Basal Ganglia neurons from patients with Parkinson's disease share mathematical features with the velocity fields of fluids under turbulence regimes. In particular the properties that are similar are related to the behavior of energy, and therefore fluid dynamics' concepts might be useful in the future for the understanding of the nervous system from a mathematical perspective. In the present work we try to build the fundaments for a physical analogy between both kinds of systems.

**Introduction**

To present date, the structure of the neural code, or even if there is one, remains unknown.[1,2] Several approaches have been proposed to solve the riddle. Different works have discussed the idea of information being coded in a rate code or a time code.[3,4] In the first case the exact time of occurrence of a single spike could be random, while in the second case a deterministic time pattern is thought to code for information. In a recent work of our group, it has been proposed that information could also be coded in a scale code, with different scales of the neuronal discharge transmitting either complementary or redundant information.[5] Of course, different combinations of the previous might be possible, and the universality of the code is also under discussion. Different neural systems might use different codes and even switch to different forms of coding under different circumstances.[6,7]

In the present work we will discuss a general frame to analyze the circulation of nervous impulses throughout the nervous system. In previous works we have shown that properties measured from the brain from patients with Parkinson's disease (PD) have similarities with turbulent fluids.[5,8,9] We have proposed that Navier-Stokes-like equations might be useful for the modeling of information transmission in the central nervous system. In the present paper we present a preliminary description of a physical analogy and a possible way in which the formalism of Navier-Stokes (NS) might apply for the study of information transmission in a neural system.

A brief discussion about the possible consequences of the present work might be appropriate. In clinical neurology the question about the way in which neurons code information becomes relevant for the understanding and modeling of information transmission in the central nervous system. In particular in the case of Parkinson's disease, the application of Deep Brain Stimulation therapy (DBS) affects the transmission of nervous impulses through the basal ganglia-thalamic-cortical (BG-Th-Ctx) loop.[10,11] DBS consists of chronically implanting a stimulation electrode in a chosen neural center and it has proven to succeed in the treatment of different conditions.[12] However, to present date the mechanisms of action of DBS therapy are not fully understood.[13,14] This poor understanding of the exact mechanisms of action of DBS has important practical consequences. Today, there exist limitations in the outcome prediction of patients with PD who are treated with DBS, and the indications of this therapy are limited to the most severe cases.[15] Therefore the fact that we currently lack a mathematical model of DBS poses important limitations to this treatment. Based on some experimental evidence presented in previous works [5,8,9] we will discuss a possible analogy between the circulation of spikes in the BG-Th-Ctx loop and the dynamics of fluids.

**Fundaments of the model**

The present model is centered on the description of the velocity of transmission of the nervous impulse in the nervous system. Although velocity is frequently not used to describe the behavior of neural systems, it might possibly be a crucial variable for the understanding of information transmission.[16] To understand the situation that we are trying to represent, imagine a simple neural system like the one in figure 1. Suppose that a spatially homogeneous stimulus S1 is applied to three adjacent neurons, N1-3, forming layer one (LI) of a neural network. All three neurons receive the stimulus at the same time, called $t_i$, and we propose that the three of them will respond with the generation of a single spike to the stimulus. The problem that concerns us is at what times, called $t_1$, $t_2$ and $t_3$, the spikes originated by S1 will arrive to the next layer of the network travelling through the different neurons. (For simplification we will consider here that the next layer lies at a constant distance from layer one, although the spatial architecture of physiological neural systems adds a complexity level to the problem that we are analyzing. This fact could be modeled with the use of non-Bravais lattices, i.e. integration meshes without translational symmetry.) Assuming that spikes travel at the same velocity through all the axons of the layer, $t_1$, $t_2$ and $t_3$ will depend on the time consumed in summation and therefore on the level of excitation of each neuron. We assume that all the cells in a layer have the same time and space constants and firing threshold. Therefore changes in membrane potential will occur at the same rate in each cell when the depolarizing stimulus is applied and, if the stimulus is enough to produce a supra-threshold depolarization, then the time at which the response will be generated depends on how far the membrane potential lies from the threshold at the moment the cell received the input. It can be easily seen that this difference in the initial excitation level of each neuron will translate into a difference between $t_1$, $t_2$ and $t_3$, a phenomenon that can actually be interpreted as a change of velocity of the impulse if we consider the network globally. This acceleration effect could be potentiated or attenuated in the next layer, adding variability to the velocity of transmission of spikes in the nervous system and finally generating the complex interspike intervals time series that are usually measured in neural systems.

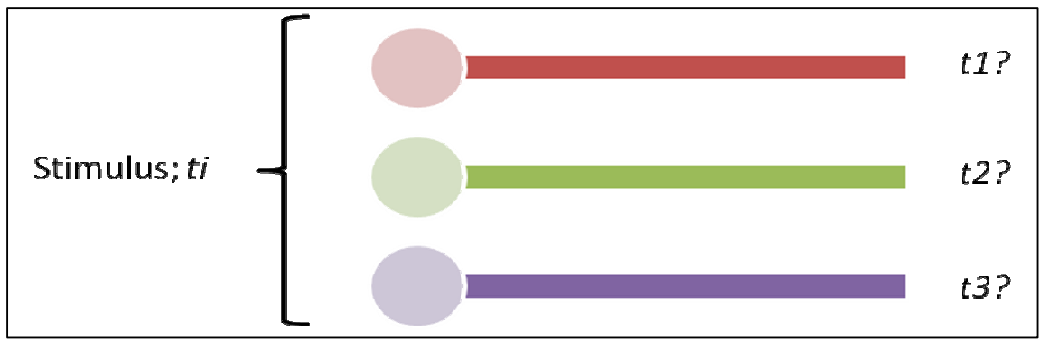

Figure 1) Although the velocity of spikes is a quantity that is not commonly discussed, it might well describe the circulation of impulses in the nervous system. Suppose that three adjacent neurons are affected simultaneously by a spatially homogeneous stimulus at an initial time *t*, and assume that all three neurons have the same time and space constants and firing threshold and a constant axonal length. If the three neurons respond with the generation of a single spike to the stimulus, then the times *t1, t2* and *t3* at which the generated spikes will arrive to the next neuronal layer do not need to be the same, but will depend on the exact level of excitation at of each neuron at time *t*.

So far we have described a simple fact about neural transmission that is often not taken into account. This is also true for another simple fact that needs to be considered prior to the discussion of the present model. Given two neurons N1 and N2, any of them will have a greater probability of applying a positive acceleration to an impulse $I$ if the other one is doing so. This is equivalent to saying that the velocity through a given neuron will be minor if the adjacent neuron is not applying any acceleration to the impulse. This can be explained with our current knowledge of extracellular modulation of synaptic activity, a function conducted primarily by astrocytes. [17-19] The importance of the finite extracellular space and its chemical composition in neural conduction is being acknowledged progressively, and is a possible mechanism of coupling between adjacent cells. [20,21] When considering a large network of neurons this tendency of adjacent neurons to show similar acceleration values will have the effect of helping a spatially homogeneous stimulus to remain so. In other words, while acceleration and deformation in time and space of the stimulus travelling through the network is possible, the stimulus will essentially resist to this deformation. We will build an analogy between this fact and the viscosity of fluids, and will call this property of stimuli travelling through large ensembles of neurons, $\mu$.

Having stated that the velocity of transmission of an impulse from one layer to the next one is not necessarily constant and that a spatially homogeneous stimulus naturally resists deformation, we have the bases needed to apply the present NS-based model to the transport of nervous stimuli through the neural network that we are considering.

We can write the well known Navier-Stokes equations as follows:

$$\vec{g} = 0 \qquad (1)$$

$$\Rightarrow \rho \frac{\partial \vec{v}}{\partial t} = -\nabla P + \mu \nabla^2 \vec{v} \qquad (2)$$

Where $\mu$ is the viscosity, P stands for pressure and $v$ for velocity. We will consider $g = 0$ in order to ignore the gravity term. To make use of this equation for an interpretation of the movement of nervous impulses in the nervous systems, we will define $\rho$ as the density of impulses $I$ in a given volume of nervous tissue. Pressure can also be considered as the energy of the system per unit of volume. For that reason the best analogy possible in the nervous system should be a function of the excitation level of the neuron. In time, the level of excitation can be described as a function of the sum of synaptic weights in order to make comparisons with other models. Since the general state of excitation depends on the sum of synaptic weights of all the synapses present in the system, we can write excitation, $e$, as a function of this sum:

$$P = e\left(\sum_{m}^{i=1} w_i\right) \qquad (3)$$

What was said to this point allows us to state that the propagation of the nervous impulse through the nervous system requires energy per volume unit. We have already stated that $\mu$ will account for the resistance to deformation of a spatially homogeneous stimulus. These ideas are just being formulated in a conceptual way in the present work, and experimental research needs to follow in order to describe in a quantitative way these properties.

A few words need to be said in reference to a possible comparison to experimental data. Interspike intervals (ISIS) are a common, robust and reliable measure used for the analysis of experimental data in neural systems. Since the detection of spikes is a robust procedure itself, these time series have low levels of noise, making them suitable for the analysis of non-linear properties. Interspike intervals time series are constructed using the differences between the times of appearance of spikes, and are therefore a measure of time. A practical difficulty that we will encounter when trying to apply NS-based model to the circulation of impulses in the nervous system is that most works based on NS models analyze velocity fields. For that reason a way to compare time and velocity would be desirable, and some transformation between both variables might be needed. This transformation can be performed making use of single cell recordings. When analyzing single unit data, it can be considered that every spike is generated at the same point of space. This allows us to take distance as 1, and so a relationship of $x^{-1}$ relates both variables. Of course, some normalization would be needed when working with velocity fields in more than one dimension. Then the inverse relationship would be to the norm of vector velocity, and the relationship would be:

$$I(x,t) \tag{4}$$

$$ISI = t_i - t_{i+1} \tag{5}$$

$$\bar{v} = \left(\frac{\Delta x}{\Delta t}, \frac{\Delta y}{\Delta t}, \frac{\Delta z}{\Delta t}\right) \tag{6}$$

$$\Delta x_{LI-LII} = \Delta y_{LI-LII} = \Delta z_{LI-LII} = 1 \tag{7}$$

$$\Rightarrow \|v_{LI-LII}\| = \frac{\sqrt{3}}{ISI} \tag{8}$$

Here $I(x,t)$ describes the impulse as a function of time and space and $\Delta x_{LI-LII}$, $\Delta y_{LI-LII}$ and $\Delta z_{LI-LII}$ refer to the distance in three dimensional space between neuronal layers I and II (we are essentially considering a cube of side =1).

**Discussion**

During the last decades our understanding of BG physiology and pathophysiology has been challenged by the introduction of DBS therapy. [22-23] The observations made in the context of stimulation treatment of human patients with PD as well as functional neurosurgery showed many inconsistencies and paradoxes with the classical Albin-De Long's model of the BG, opening the way for new models, different than this classical "box and arrow" view.[24] There are many important points that the classical model fails to account for. First of all, it has been shown that high frequency DBS of the GPi is effective in the treatment of hyperkinetic (chorea) as well as hypokinetic (PD) disorders in exactly the same way. [25-26] This fact is contradictory with the notion that hyperkinetic and hypokinetic diseases are based on opposite pathogenic mechanisms. Secondly, stimulation of almost any nuclei of the BG-Th-Ctx loop yields positive results in the treatment of at least some PD symptoms. This includes stimulation of GPi, GPe, ventrolateral Thalamus, subthalamic nucleus, motor cortex and even zona incerta. [27-31] Finally, it has been described that HF stimulation of the GPi increases the activity of this nucleus, which is exactly the mechanism proposed as the origin of the symptoms in the classical model of PD. [32-35] In this context, new hypothesis have been made about the functioning of the normal and diseased BG. Current models include dimensionality reduction models, action-selection devices, learning reinforcement and some neural network models. [36-40] These models capture some properties of the dynamics, but up to now none of them has proven successful enough to accurately predict outcome and allow a better programming of the devices. It has also been proposed that non-linear features might be the basis of the functioning of the BG-Th-Ctx loop and that some of these properties might be altered in PD. [41-45] However, currently there is not a mathematical model that allows representing and analyzing these properties in a quantitative fashion.

In previous works we have found some interesting mathematical similarities between the ISIS time series of parkinsonian GPi neurons and velocity fields of fluids undergoing turbulence. [5,8,9] In particular, an exponential decay of the power spectrum makes it possible to talk about critical scale length, a property that is not shared by many systems in the nature. This form of decay of the spectrum is typical from turbulent fluids, and follows an analytic solution of Navier-Stokes equations when energy remains bounded. [46,47] Therefore it could be thought that energy behaves in a similar manner in both systems. Other similarities that we have reported refer to the existence of positive Lyapunov exponents [48,49] and a scaling of the structure function that is typical from turbulent fluids too [50] and that lead to the description of a multifractal behavior in the signals studied. In light of the evidence found up to this point it seems reasonable to speculate that the equations underlying the dynamics of the neural system under discussion might be similar in form to those that better describe the behavior of fluids: Navier-Stokes equations. A further step is what we have tried in the present work, and it consists of looking for a physical analogy that explains the sense in which each term of the equations might apply for neural systems.

Another type of neurophysiological model are neural fields. These are concerned with firing rates more than with velocity, [51-55] although sometimes they have been used to describe the changes of velocity present in the nervous system[16]. However they are mainly based on reaction-diffusion kinds of systems and work under the assumption of density heterogeneities, a description that lacks of a strong physiological basis. In turn these heterogeneities are usually modeled as continuous functions of space, a condition that does not seem very physiological either. So we can say confidently that up to now there is not a model that handles the behavior of the velocity of spikes in the nervous system in a quantitative and physiological enough manner.

Currently there are several experimental approaches to the study of PD and DBS. Under some experimental settings, neurons from the basal ganglia from patients with PD have shown an oscillatory behavior.[56-59] Under this evidence it has been proposed that a pathologic increment of oscillatory activity or excessive synchronization between neurons might be the pathogenic mechanism of PD.[60] However, other works have shown the mentioned neurons to behave in a complex fashion, with an irregular output.[8,43] The fundamental problem is to what extent these different works are comparable. Probably at least some of the mathematical properties of the signals studied depend on the experimental settings used, such as in vitro vs. in vivo experiments, the level of external input of the neuron, the type of anesthesia used, etc. [61-65] In the previous works by our group that we have mentioned before, we have worked with data from GPi neurons form patients with PD. This data were recorded during functional neurosurgery, with the patients awake and under local anesthesia only. Because of that reason the findings observed might be due at least to some extent to the state of consciousness of the patients, which defines the high dimensional input from the neurons studied at the times of the recordings. Since the mathematical properties analyzed were described under these particular experimental conditions, some questions follow that will need to be considered in future works. A control group of a healthy population of neurons will be needed. For the reasons that were discussed, the best control group would probably consist of recordings performed on animals awake with microelectrode implanting techniques. This experiment together with the analysis of anesthetized animals could help elucidate if the observations under discussion are due to the presence of PD itself or to the state of consciousness of the animal and to what extent.

Of course, the hypothesis presented here need to be tested. Many directions of research follow from the idea that NS-based models might be useful for the modeling of the nervous system. First of all, an analytic work might be useful to test the relation between ISIS and velocity fields, i.e. to test if the results already known for velocity fields and the properties observed for ISIS time series hold under an $x^{-1}$ transformation. Secondly, to perform simulations for simple conditions might be useful, obtaining a comparable measure from them (times between events). For example, tissue infarctations from different sizes and shapes might be a good starting point to run simulations (different geometric boundary conditions). To model the effect of DBS it would be probably necessary to work with the pressure term of the equations. Since DBS stimulates the nervous tissue with a pulse of electric tension, it can be interpreted that it produces a change in the level of energy of the tissue (by affecting ionic gradients for example), and an analogy could be easily built with the pressure of fluids. Finally, new experimental approaches could be design to study the movement of impulses in nervous tissue from the perspective of velocity. However, and although new experiments might be desirable, it needs to be considered that experimental complexity needs to be kept to a minimum when working with human patients. For that reasons new experiments might be necessarily conducted in laboratory animals, or even more brain slices or cell cultures and, again, comparison with data from human brains will become a problem to be considered. Finally, if the presented approach turns out to be useful for the modeling of the BG, the universality of the model will become a question, and other neural systems under different conditions (awake, anesthetized, *in vivo* vs. *in vitro*) will need to be tested.

## Conclusion

In the present work we do not present simulations or computational results. We have explained the fundamentals of a possible application of Navier-Stokes-based equations for the modeling of the movement of nervous impulse in the nervous system. Velocity and momentum of spikes are not usually considered when modeling information transmission in the nervous system, but recently we have found evidence that interspike intervals time series share some mathematical properties with the behavior of velocity fields of fluids under turbulence. In particular these properties are related to the behavior of energy, and therefore fluid dynamics' concepts might be useful in the future for the understanding of the nervous system, both from a mathematical perspective and helping us to find new physical analogies.